\documentstyle[preprint,aps,eqsecnum,amsfonts]{revtex}
\begin{document}

\draft

{\tighten
\preprint{\vbox{\hbox{WIS-94/16/Mar-PH}
                \hbox{hep-ph/9403376}
                \hbox{March 1994} }}

\title{The Inclusive $\bar B\to\tau\,\bar\nu\,X$ Decay in \\
       Two Higgs Doublet Models}

\author{Yuval Grossman and Zoltan Ligeti}

\address{Department of Particle Physics \\
         Weizmann Institute of Science \\
         Rehovot 76100, Israel}

\maketitle

\begin{abstract}
We calculate in the framework of two Higgs doublet models the
differential decay rate for inclusive $\bar B\to\tau\,\bar\nu\,X$
transitions to order $1/m_b^2$ in the heavy quark expansion, for
both polarized and unpolarized tau leptons.  In contrast to the
situation in the standard model, we find sizeable $1/m_b^2$ corrections.
A systematic heavy quark expansion significantly reduces the theoretical
uncertainties in the calculation compared to the existing free quark
decay model results.  From the experimental measurement of the branching
ratio, we derive the bound $\tan\beta < 0.51\, m_{H^\pm}\,[{\rm GeV}]$.
We point out that the tau polarization is potentially a more sensitive
probe of multi scalar models than the branching ratio itself.
\end{abstract}
}%end tighten

\newpage
\narrowtext

\def\({[}
\def\){]}
\def\r{\rho}
\def\rt{\rho_\tau}
\def\Mt{\hat m_\tau}
\def\lo{\lambda_1}
\def\lt{\lambda_2}

\section{Introduction}

Semileptonic $B$ meson decays into a tau lepton are sensitive to
extensions of the standard model (SM) with several Higgs fields.
In two Higgs doublet models (2HDM) \cite{hhg} the experimental
measurement of the decay rate $\bar B\to\tau\,\bar\nu\,X$ provides
the strongest upper bound on $\tan\beta/m_{H^\pm}$
\cite{KrPo,Kali,Isid,Yuval}, for large values of $\tan\beta$.
Theoretically, charged scalar contributions could increase
$BR\,(\bar B\to\tau\,\bar\nu\,X)$ by as much as an order of
magnitude compared to the SM \cite{KrPo,Kali,GrHo}.
This could have provided an exciting solution to the semileptonic $B$
meson branching ratio problem \cite{AP,BBSV}.  While this scenario is
ruled out by the recent ALEPH measurement \cite{Aleph}, the ratio of
decay rates into the tau and light lepton channels still provides a
sensitive probe to multi scalar models.  Moreover, the possibility to
study the tau polarization offers a greater variety in both the
experimental and the theoretical analyses\cite{Kali}.

Recently, it has been observed that inclusive semileptonic decays of
hadrons containing a single heavy quark allow for a systematic,
QCD-based expansion in powers of $1/m_Q$ \cite{CGG}.
The heavy quark limit ($m_Q\to\infty$) coincides with the free quark
decay model and there are no corrections to this result at order
$1/m_Q$ \cite{CGG,BUV}.
The leading nonperturbative corrections are of order $1/m_Q^2$ and
depend on only two hadronic parameters, which parameterize forward
matrix elements of local dimension-five operators in the heavy quark
effective theory (HQET) \cite{review}.
These corrections have been computed for a number of
processes \cite{BUV,Bigi,MaWi,FLS,Mann,else,us}.

In the framework of the SM it was shown \cite{us} that the
heavy quark expansion provides more precise predictions than the
spectator quark model for inclusive $\bar B\to\tau\,\bar\nu\,X$ decays.
The effect of including the nonperturbative $1/m_Q^2$ corrections
is numerically small.  More important is that HQET gives an unambiguous
definition of the quark masses \cite{FNL} that determine the inclusive
decay rates.  Using the relation between the masses of the heavy quarks
in terms of the HQET parameters leads to a significant reduction in the
theoretical uncertainties \cite{us}.

In this paper, we use a combination of the operator product expansion
and HQET to study the decay $\bar B\to\tau\,\bar\nu\,X$ in the
framework of 2HDM.  In section 2 we define the model.  In section 3,
we present the analytic results to order $1/m_b^2$ for the branching
ratio and for the tau polarization.  In section 4, we study the
numerical predictions for the total decay rate and for the tau
polarization.  Section 5 contains our conclusions.

\section{The model}

Extensions of the standard model scalar sector are constrained by the
experimental value of $\rho\equiv m_W^2/(m_Z^2\cos^2\theta_W)\simeq1$
\cite{PDG} and by the strong limits on flavor changing neutral currents
(FCNC).  The first of these constraints is naturally fulfilled if the
Higgs sector contains only doublets (and singlets), providing the tree
level prediction $\rho=1$.  FCNC are eliminated at tree level if we
assume natural flavor conservation (NFC) \cite{GW}, a construction that
requires fermions of a given electric charge to couple to a single Higgs
doublet.  The phenomenology of such models has been widely studied
\cite{hhg}.  In these models weak decays are mediated besides the
$W^\pm$ bosons by charged scalar particles as well.

For simplicity we consider 2HDM with NFC and comment on the more
general case below.  After spontaneous symmetry breaking only two
physical charged scalars remain, $H^\pm$.
Such a model has still several variants.
If the Higgs doublet that couples to up-type quarks also couples to
down-type quarks or to charged leptons then the experimental results on
the decays $B\to X_s\gamma$ \cite{Hewett} and $Z\to \bar b b$ \cite{Park}
yield very strong constraints and the scalar contribution to
$\bar B\to\tau\,\bar\nu\,X$ is negligible.
Therefore, we only consider models in which one Higgs doublet couples
to down-type quarks and charged leptons and the other to up-type quarks
(model $I\!I$ in the language of \cite{hhg}).
The minimal supersymmetric standard model belongs to this class,
so our results also apply to this model (up to radiative corrections).

The Yukawa interaction of the physical charged scalars with fermions is
determined by $\tan\beta$ (the ratio of the vacuum expectation values
of the two Higgs doublets), by the fermion masses and by the CKM matrix.
The terms in the effective lagrangian relevant for
$\bar B\to\tau\,\bar\nu\,X$ decays are
\begin{equation}\label{lag}
{\cal L}= -V_{cb}{4G_F\over\sqrt2} \Big\( (\bar c\gamma^\mu P_L b)\,
(\bar\tau\gamma_\mu P_L\nu_\tau) - R\, (\bar c P_R b)\,
(\bar\tau P_L\nu_\tau) \Big\) \,,
\end{equation}
where
\begin{equation}\label{Rdef}
R = r^2\, m_\tau\, m_b^Y \,, \qquad
r={\tan\beta\over m_{H^\pm}} \,,
\end{equation}
and $P_{R,L}=\frac12(1\pm\gamma_5)$.  We denoted by an upper index $Y$
the running quark masses (this distinction is unimportant for the tau,
see discussion in section 4).  The first term gives the SM contribution,
while the second one gives that of the charged scalars.
We neglect a term proportional to $m_c^Y$:
first, it is suppressed by the mass ratio $m_c^Y/m_b^Y$; second,
it cannot be enhanced by the possibly large factor $\tan^2\beta$.

Our analysis holds with minor modifications for general multi Higgs
doublet models with NFC (for a recent comprehensive analysis, see
\cite{Yuval}).  In such models, instead of the single parameter
$\tan\beta$, three complex coupling constants determine the Yukawa
interactions of the charged scalars.  The parameters $X$, $Y$ and $Z$
describe the couplings to up-type quarks, down-type quarks and charged
leptons, respectively.  Then in the effective lagrangian (\ref{lag})
$r^2=XZ^*/ m_H^2$, where $m_H$ is the lightest charged scalar mass
(assuming that the heavier charged scalars effectively decouple from
the fermions).  Thus, although $r^2$ is real in 2HDM, we treat it as
complex so that our results apply to general multi Higgs doublet models.

\section{Heavy quark expansion}

In this section we present the analytic results necessary for our
numerical analysis.  The techniques by which they were obtained are
described in detail elsewhere \cite{Bigi,MaWi,FLS,Mann}, so we give
only a brief summary, followed by the results of the computation.

The inclusive differential decay distribution is determined by the
imaginary part of the time-ordered product of two flavor-changing
currents,
\begin{equation}
T = -i\int d^4x\,e^{-iq\cdot x}\,\langle B|\,
   T\left\{{\cal O}^\dagger(x),{\cal O}(0)\right\} |B\rangle\,,\qquad
{\cal O} =\bar c\left(\gamma^\mu P_L - m_b^Y r P_R \right) b\,.
\end{equation}
Since over most of the Dalitz plot the energy release is large
(of order $m_b$), the time-ordered product can be written as an
operator product expansion, in which higher-dimension operators are
suppressed by powers of $\Lambda/m_b$, where $\Lambda$ is a typical
low energy scale of the strong interactions.
To this end it is necessary to separate the large part of the $b$ quark
momentum by writing $p_b=m_b v+k$, where $v$ is the velocity of the
decaying $B$ meson, and its residual momentum $k$ is of order $\Lambda$.
This separation is most conveniently performed by using the formalism of
the heavy quark effective theory \cite{review}.  One can then evaluate
the matrix elements of the resulting tower of nonrenormalizable operators
with the help of the heavy quark symmetries.

The leading term in the expansion reproduces the result of the free
quark decay model \cite{CGG}, while giving an unambiguous definition
of the heavy quark mass \cite{FNL}.  The leading nonperturbative
corrections are of relative order $1/m_b^2$ and can be written in terms
of two parameters, $\lambda_1$ and $\lambda_2$, which are related to the
kinetic energy $K_b$ of the $b$ quark inside the $B$ meson, and to the
mass splitting between $B$ and $B^*$ mesons:
\begin{equation}\label{lambdadef}
K_b = - {\lambda_1\over 2m_b}\,,\qquad
m_{B^*}^2 - m_B^2 = 4\lambda_2\,.
\end{equation}

The operator product expansion for the SM contribution has been
presented in refs.\cite{else,us}.  Including the charged Higgs
contribution is, in principle, a straightforward generalization.
We shall present only the final results.
The tau lepton can have spin up ($s=+$) or spin down ($s=-$) relative
to the direction of its momentum, and it is convenient to decompose the
corresponding decay rates as
\begin{equation}
\Gamma\left(\bar B\to\tau(s=\pm)\,\bar\nu\,X\right)=
    {1\over2}\Gamma\pm\tilde\Gamma\,.
\end{equation}
The total rate, summed over the tau polarizations, is given by $\Gamma$,
while the tau polarization is $A_{\rm pol}=2\tilde\Gamma/ \Gamma$.
Let us further decompose
\begin{equation}
\Gamma= {|V_{cb}|^2\, G_F^2\, m_b^5 \over 192\pi^3}
\left\( \Gamma_W + {|R|^2\over4}\,\Gamma_H
-2\,{\rm Re}(R)\,{m_\tau\over m_b}\,\Gamma_I \right\)\,, \qquad
\end{equation}
and similarly for $\tilde\Gamma$. The subindices $W$, $H$, and $I$
denote the $W$ mediated (standard model), Higgs mediated and
interference contributions, respectively.
This notation is particularly convenient, because in the spectator
model $\Gamma_W =\Gamma_H$, $\tilde\Gamma_H =-\tilde\Gamma_W$
and $\tilde\Gamma_I=0$.
In the $m_\tau,\,m_c\to 0$ limit of the spectator model
$\Gamma_W =\Gamma_H =\Gamma_I =2\tilde\Gamma_H =-2\tilde\Gamma_W =1$.
After integration over the invariant mass of the lepton pair and the
neutrino energy, the differential decay rate depends only on a single
kinematic variable $E_\tau$, which denotes the tau energy in the rest
frame of the decaying $B$ meson.  We will use the dimensionless variables
\begin{equation}
y = {2E_\tau\over m_b}\,,\qquad
\rho = {m_c^2\over m_b^2}\,,\qquad
\rho_\tau={m_\tau^2\over m_b^2}\,.
\end{equation}

The standard model contribution to the lepton spectrum
($d\Gamma_W/dy$ and $d\tilde\Gamma_W/dy$) has been calculated in
ref.\cite{us}, and we do not present it here.
In terms of these quantities, however, the Higgs mediated
contribution is quite simple.  The reason is that the kinetic operator
does not violate the heavy quark spin symmetry, thus the corrections
proportional to $\lo$ are the same for the $W$ and Higgs mediated terms.
We obtain
\begin{eqnarray}
{d\Gamma_H\over dy} &=& {d\Gamma_W\over dy} +{\lt\over m_b^2}
{24x_0(1-x_0)\(y(3-y) -\rt(8-3y)\)\sqrt{y^2-4\rt}\over 1+\rt-y} \,,
\\[12pt]
{d\tilde\Gamma_H\over dy} &=& -{d\tilde\Gamma_W\over dy}
+{\lt\over m_b^2} {12x_0(1-x_0)\,(y^2-4\rt)(3-y-\rt)\over 1+\rt-y} \,,
\end{eqnarray}
where
\begin{equation}
x_0=1-{\rho\over1+\rt-y}\,.
\end{equation}
For the interference term we get:
\begin{eqnarray}\label{GI}
{d\Gamma_I\over dy} &=& \sqrt{y^2-4\rt}\, \bigg\{
    6x_0^2(1+\rt-y)\nonumber\\
    &-& {\lo\over m_b^2(1+\rt-y)} \Big\(
    2(y^2-4\rt) + 2x_0\left(y^2-3y(1+\rt)+8\rt\right) \nonumber\\
    &&\qquad\qquad\qquad\quad -x_0^2
    \left(3+14\rt+3\rt^2-9y(1+\rt)+4y^2\right) \Big\)\nonumber\\
    &-& {3\lt\over m_b^2(1+\rt-y)}\Big\(
    2x_0\left(12+8\rt-3y(5+\rt)+4y^2\right) \nonumber\\
    &&\qquad\qquad\qquad\quad -x_0^2
    \left(15+10\rt+3\rt^2-3y(7+3\rt)+8y^2\right)\Big\)\bigg\}\,,
\end{eqnarray}
and
\begin{equation}\label{GtI}
{d\tilde\Gamma_I\over dy}=
-{\lt\over m_b^2} {3x_0(1-x_0)\,(y^2-4\rt)(2-y)\over 1+\rt-y} \,.
\end{equation}

The explicit form of the total decay rates are obtained by integrating
the above expressions over the kinematic range
$2\sqrt{\rt} \leq y \leq 1+\rt-\r$.
We do not reproduce here the explicit expressions for the standard model
contributions $\Gamma_W$ and $\tilde\Gamma_W$ that can be found in
ref.\cite{us}.  For the Higgs mediated term we obtain:
\begin{eqnarray}
\Gamma_H &=& \Gamma_W + {12\lt\over m_b^2}
\bigg\{ \sqrt\lambda\, \(2(1-\rt)^2+\r(5-\r+5\rt)\)
-6\r(1-\rt^2)A -6\r(1+\rt^2)B \bigg\}\,, \\[12pt]
\tilde\Gamma_H &=& -\tilde\Gamma_W +
{6\lt\over m_b^2} \bigg\{ \((1-\Mt)^2-\r\)\,
\Big\(2(1-\rt)(1+\Mt)^2-\r^2 \nonumber\\
&&\quad\qquad\qquad\qquad\qquad\qquad\qquad
+\r(5-3\Mt-\rt+15\rt\Mt)/(1-\Mt)\Big\) \nonumber\\
&&\qquad\qquad\quad -2\r(3+2\rt-4\rt\r-5\rt^2) \ln{(1-\Mt)^2\over\r} \bigg\}\,,
\end{eqnarray}
where we defined
\begin{equation}
A= \ln{2\Mt\r\over(1-\rt)^2-\r(1+\rt)-(1-\rt)\sqrt\lambda}\,, \qquad
B= \ln{2\Mt\over1+\rt-\r+\sqrt\lambda}\,,
\end{equation}
$\lambda=1-2(\rt+\r)+(\rt-\r)^2$, and $\Mt=m_\tau/m_b=\sqrt{\rho_\tau}$.
For the contribution of the interference term we obtain
\begin{eqnarray}
\Gamma_I =&& \sqrt\lambda\bigg\{\Big(1+{\lo\over2m_b^2}\Big)
(1-5\r-2\r^2+10\rt-5\rt\r+\rt^2) \nonumber\\
&&\qquad -{3\lt\over2m_b^2} (11-\r+2\rt+8\r^2+35\rt\r-\rt^2) \bigg\}\nonumber\\
+ && 6\Big(1+{\lo\over2m_b^2}\Big)
\Big\(\r^2(1-\rt)A +\Big((\r^2+2\rt)(1+\rt)-4\rt\r\Big)B\Big\)\nonumber\\
+ && {9\lt\over m_b^2} \Big\(\r(2+\r-5\rt\r-2\rt^2)A
+\Big(2\r(1-\rt)^2-6\rt+2\rt^2+\r^2+5\rt\r^2\Big)B \Big\)\,,
\end{eqnarray}
and
\begin{eqnarray}
\tilde\Gamma_I = -{3\lt\over 2m_b^2} &\bigg\{& \((1-\Mt)^2-\r\)\,
\Big\((1-\rt)(1+\Mt)^2-\r^2-2\Mt\r(1-5\rt)/(1-\Mt)\Big\) \nonumber\\
&-& 2\r(1-\r-\rt)(1+3\rt) \ln{(1-\Mt)^2\over\r} \bigg\}\,.
\end{eqnarray}
In the limit $\lambda_1,\lambda_2\to0$, corresponding to the free quark
decay model, our results agree with \cite{Kali}.  However, we disagree
with \cite{Isid} on the size of the interference term.  There it was
presumably multiplied with an extra factor of $m_\tau/m_b$.  As expected,
the parameter $\lambda_1$ only enters the total decay rate proportional
to the spectator model result, namely the total decay rate depends on
$\lambda_1$ only through the combination $(1+\lambda_1/2m_b^2)$.
% While the total rate in the SM is symmetric in $\r$ and $\rt$, this is
% not true for the Higgs and the interference terms.
By virtue of the coupling of the charged scalar to the lepton pair
(\ref{lag}), the Higgs mediated term produces tau leptons with
positive polarization.  This indicates that, in addition to the decay
rate, the tau polarization is also sensitive to new physics.

\section{Numerical results}

\subsubsection{Input parameters}

When we neglect the tiny contribution from $b\to u$ transitions, the
parameters entering our calculations are the mass of the tau lepton
$m_\tau = 1.777\,{\rm GeV}$ \cite{mtauexp}, the heavy quark masses
$m_b$ and $m_c$, the hadronic parameters $\lambda_1$ and $\lambda_2$,
the quark mixing parameter $|V_{cb}|$, and the combination of the
unknown Higgs sector parameters $R$ (\ref{Rdef}).
To utilize the full power of the heavy quark expansion it is important
that not all of these parameters are independent \cite{us,YoZo}.
In fact, the same HQET parameters $\lambda_1$ and $\lambda_2$ appear
in the expansion of the heavy meson masses in terms of the charm and
bottom quark masses (up to the running of $\lambda_2$)\cite{FaNe}:
\begin{eqnarray}\label{cbmasses}
m_B &=&m_b+\bar\Lambda-{\lambda_1+3\lambda_2\over2m_b}+\ldots\,,
\nonumber\\
m_D &=&m_c+\bar\Lambda-{\lambda_1+3\lambda_2\over2m_c}+\ldots\,.
\end{eqnarray}
The parameter $\bar\Lambda$ can be associated with the effective mass
of the light degrees of freedom inside the heavy
meson \cite{Luke,FNL}\footnote{%
It has been argued recently \cite{renormalon} that the definition
of $\bar\Lambda$ is ambiguous.  If this ambiguity turns out to be
numerically significant that could probably be accounted for by taking
a larger range for $\bar\Lambda$ in our numerical calculations and would
slightly increase the uncertainties.}.
For each set of values $\{\bar\Lambda,\lambda_1,\lambda_2\}$,
eq.~(\ref{cbmasses}) determine $m_c$ and $m_b$.
The consistency of the heavy quark expansion requires that these values
of the quark masses are used in the theoretical expressions for the decay
rates.  Using only three independent parameters
$\{\bar\Lambda,\lambda_1,\lambda_2\}$ instead of the four quantities
$\{m_b,m_c,\lambda_1,\lambda_2\}$ reduces the theoretical uncertainties
significantly.

These HQET parameters are genuinely nonperturbative.  While the
value of $\lambda_2$ is related to the $B^*-B$ mass splitting
$\lambda_2 =(m_{B^*}^2-m_B^2)/4 \simeq 0.12\,{\rm GeV}^2$,
there is no similarly simple way to determine $\bar\Lambda$ and
$\lambda_1$.  We expect this value of $\lambda_2$ to be accurate
to within 10\%, as a result of the finite $b$ quark mass and the
experimental uncertainties.  The parameters $\bar\Lambda$ and
$\lambda_1$ can only be estimated in models of QCD at present, or
constrained from phenomenology\cite{YoZo}.
QCD sum rules have been used to compute $\lambda_1$ \cite{SRNeu,SRlam},
but these calculations suffer from large uncertainties.
There is increasing theoretical evidence that, in accordance to its
definition (\ref{lambdadef}), $\lambda_1$ is negative \cite{lambdaneg},
and its magnitude cannot be too large \cite{virial}.
Here we use $0< -\lambda_1 <0.3\,{\rm GeV}^2$, which
is also supported by a recent QCD sum rule calculation \cite{Neubert}.
Assuming $\lambda_1<0$, the phenomenological analysis of ref.\cite{YoZo}
prefers values of $\bar\Lambda$ lower than QCD sum rules
\cite{SRNeu,SRLam}.  Here we take $0.4< \bar\Lambda <0.6\,{\rm GeV}$.

Finally, a subtle point is that of the perturbative QCD corrections.
While we know exactly the ${\cal O}(\alpha_s)$ corrections to the
$W$ mediated term in the spectator model, analogous calculations
for the $H$ and $I$ terms have not been carried out.  For the $W$
term $\eta_\tau\simeq 0.90$ \cite{HoPa}; we shall also use the
${\cal O}(\alpha_s)$ correction to $\bar B\to e\,\bar\nu\,X\,$:
$\eta_e\simeq 0.88$.  The leading corrections to the Higgs coupling are
incorporated by running the Yukawa coupling.
Therefore, $R$ depends on the running mass of the $b$ quark $m_b(\mu)$.
We take a conservative range $0.9\,m_b\leq m_b^Y \leq 1.05\,m_b$
corresponding to $m_c\leq\mu\leq m_b$.
For the interference term we include an extra $(-5\pm10)$\,\%
correction to represent the uncertainty in the QCD correction.

To conclude the above discussion we summarize the ranges for the
various input parameters that we use in our analysis:
\begin{equation}\label{ranges}
   \begin{array}{ll}
0.4< \bar\Lambda <0.6\,{\rm GeV} \,, &\qquad
0< -\lambda_1 <0.3\,{\rm GeV}^2 \,, \\
0.9\,m_b < m_b^Y < 1.05\,m_b \,, &\qquad
0.11< \lambda_2(m_b) < 0.13\,{\rm GeV}^2 \,.
   \end{array}
\end{equation}

\subsubsection{Branching ratio}

Normalizing the branching ratio in the tau channel to that into light
leptons separates the theoretical and experimental uncertainties in the
numerical analysis, eliminates the otherwise significant uncertainties
from the values of $|V_{cb}|^2$ and $m_b^5$, and also reduces the
sensitivity to the unknown QCD corrections.
We plot the theoretical prediction for the branching ratio as a
function of $r$ in Fig.~1.  Our result is given by the shaded region
between the solid lines.

Using the recent measurement of the branching fraction into final states
with a tau lepton \cite{Aleph} and that with an electron \cite{PDG},
\begin{eqnarray}
BR\,(\bar B\to \tau\,\bar\nu\,X) &=& 2.76\pm0.63\,\%\,, \nonumber\\
BR\,(\bar B\to e\,\bar\nu\,X) &=& 10.7\pm0.5 \,\%\,,
\end{eqnarray}
we obtain the $1\sigma$ upper bound
\begin{equation}
r < 0.51\,{\rm GeV}^{-1}\,.
\end{equation}
The $95$\% CL upper bound is $r<0.55\,{\rm GeV}^{-1}$.

There are a number of points to be made regarding this result:

$a$.  The $1/m_b^2$ corrections to the free quark decay model turn out to
be more significant than in the SM: while the $W$ mediated contribution
is suppressed by $-4$\% and $-8$\% in the electron and tau channels
respectively, the Higgs mediated term is enhanced by approximately
14\%, and the interference term is suppressed by as much as 25\%.
As a result, the difference between the 2HDM and the SM becomes more
significant than expected from the free quark decay model.

$b$.  The uncertainty in our result (corresponding to the width of the
shaded region in Fig.~1) takes into account the theoretical
uncertainties in (\ref{ranges}), as well as an estimate of the
$1/m_Q^3$ and more importantly the perturbative QCD corrections.
If the latter terms were known, the width of the shaded region would be
reduced by about a factor of two for not very small values of $r$.

$c$.  The improvement of our calculation over the spectator model
originates from large $1/m_b^2$ corrections, and from using $m_b$ and
$m_c$ as determined by eq.~(\ref{cbmasses}) rather than treating them
as independent input parameters.
To illustrate this, we plotted with dashed lines in Fig.~1 the
prediction of the free quark decay model corresponding to
$1.4<m_c<1.5\,{\rm GeV}$ and $4.6<m_b<5\,{\rm GeV}$.
Our predictions are not very sensitive to the values of $\bar\Lambda$
and $\lambda_1$.  Adopting the range $0.3<\bar\Lambda<0.7\,{\rm GeV}$
and $-0.5<\lambda_1<0.5\,{\rm GeV}^2$, which covers a range for $m_c$
and $m_b$ larger than the ones previously mentioned, makes the allowed
band slightly wider than in Fig.~1.

$d$.  As the interference term gives rise to a dip in the branching
ratio for small values of $r$, this measurement cannot probe values of
$r$ smaller than $0.43\,{\rm GeV}^{-1}$.
Within the spectator model, this limit is $r<0.6\,{\rm GeV}^{-1}$.

$e$.  In addition to our improvement of the theoretical calculation
we would like to point out that some of the earlier analyses quoted too
strong bounds on $r$ by (a) neglecting the uncertainties in the quark
masses; (b) underestimating the size of the interference term by
about a factor of three based on \cite{Isid}; and (c) overestimating
the SM prediction for the branching ratio.

\subsubsection{tau polarization}

The polarization of the tau lepton, $A_{\rm pol}=2\tilde\Gamma/\Gamma$,
being a ratio of decay rates, is subject to much smaller uncertainties
than the rates themselves.  We find that the numerical value for
$A_{\rm pol}$ is rather insensitive to variations in $\bar\Lambda$ and
$\lambda_1$.  Allowing these parameters to vary within the ranges
(\ref{ranges}), we find the allowed values of the tau polarization as a
function of $r$.  Our result is given by the shaded region
between the solid lines in Fig.~2.

A few points are in order regarding this result:

$a$.  The $1/m_b^2$ corrections to the free quark decay model increase
$A_{\rm pol}$ for small values of $r$.  More importantly, these
corrections make the tau polarization almost a monotonically increasing
function of $r$.  While the interference term contribution to
$\tilde\Gamma$ is small (it is exactly zero in the spectator model),
it is important that the contribution of the interference term to
$\Gamma$ is significantly suppressed while that of the Higgs
mediated term is enhanced due to the $1/m_b^2$ corrections.

$b$.  The uncertainty in our result (corresponding to the width of the
shaded region in Fig.~2) takes into account the same theoretical
uncertainties as in Fig.~1.  The most important of these are the
perturbative QCD corrections: if they were known, that would make the
prediction for $A_{\rm pol}$ accurate at the 1\% level for
$r\leq0.3\,{\rm GeV}^{-1}$.

$c$.  Besides the suppression of the interference term and the
enhancement of the Higgs contribution to the total rate, the improvement
of our calculation over the spectator model is again due to using $m_b$
and $m_c$ as determined by eq.~(\ref{cbmasses}) rather than treating
them as independent input parameters.  For comparison, we plot
(with dashed lines) the prediction of the free quark decay model
corresponding to the same ranges of the quark masses as in Fig.~1.

$d$.  Most important is that a tau polarization measurement of an
accuracy of about 10\% would be sensitive to values of $r$ as small as
$0.3\,{\rm GeV}^{-1}$.  As pointed out in the previous section, such a
small value of $r$ cannot be probed with the total branching ratio.
The theoretical prediction for $A_{\rm pol}$ is accurate at the
1\% level in the standard model
$-0.70 \leq A_{\rm pol}^{\rm SM} \leq -0.71$ \cite{us}.
A precise measurement of this quantity would be a very stringent test
of the SM, or alternatively provide a determination of the parameter $r$.

Finally we would like to compare the bounds that can be achieved from
the inclusive decay $\bar B\to\tau\,\bar\nu\,X$ to that from the
exclusive decays $\bar B\to\ell\,\bar\nu$.  In ref.\cite{Hou} it
was shown that the purely leptonic decays of $B$ mesons are enhanced
compared to the SM for $r>0.27\,{\rm GeV}^{-1}$, and a bound
$r\leq0.52\,{\rm GeV}^{-1}$ has also been claimed.
It is important to note that deriving a bound on $r$
from this decay requires an independent estimate of $|V_{ub}|$ and
the $B$ meson decay constant $f_B$.  Uncertainties related to the
values of these parameters have been neglected in \cite{Hou}.
These uncertainties are, however, very significant; taking them into
account ($f_B>140\,{\rm MeV}$ and $|V_{ub}|>0.0024$)
allows one to obtain only $r<0.80\,{\rm GeV}^{-1}$.
Concerning future bounds from this process we estimate that an
enhancement of $BR\,(\bar B\to\ell\,\bar\nu)$ will only be observable
if $r\geq0.35\,{\rm GeV}^{-1}$.
We conclude that a stronger bound on $r$ will be provided
by a measurement of the tau polarization in the inclusive
$\bar B\to\tau\,\bar\nu\,X$ decay, while at present the best bound
comes from the measurement of the branching ratio in the same decay.

\section{Summary}

We investigated in detail the effects of a charged scalar on the
inclusive decay $\bar B\to\tau\,\bar\nu\,X$.
We used the heavy quark expansion to incorporate the nonperturbative
$1/m_b^2$ corrections, and the relation between the heavy quark masses
implied by the HQET.  This resulted in a reduction of the theoretical
uncertainties by more than a factor of two compared to the calculation
in the free quark decay model.  While the $1/m_b^2$ corrections are
usually less than 5--10\% in standard model calculations, in the present
case they turned out to be more significant.
To derive a bound on $r$, special care has to be taken of the
uncertainties in the theoretical calculation which have been mostly
neglected so far.
{}From the presently available experimental data the strongest bound that
can be obtained is $r=\tan\beta/m_{H^\pm}<0.51\,{\rm GeV}^{-1}$.
We pointed out that the tau polarization is more sensitive to small
values of $r$ than the branching ratio.
While $BR\,(\bar B\to\tau\,\bar\nu\,X)$ is insensitive to
$r\leq0.4\,{\rm GeV}^{-1}$, a measurement of $A_{\rm pol}$ could probe
$r\sim0.3\,{\rm GeV}^{-1}$ or even lower.
This is particularly interesting since the recent ALEPH measurement of
the branching ratio excludes very large deviations from the standard
model.  We hope that this will encourage experimentalists to measure
the tau polarization in $B$ decays with the best achievable accuracy.

\acknowledgments
We are indebted to Yossi Nir for numerous discussions and comments on
the manuscript.  It is a pleasure to thank Adam Falk, Miriam Leurer,
Matthias Neubert and Adam Schwimmer for helpful conversations.

{\tighten

\begin{figure}
\caption[a]{
$\Gamma\,(\bar B\to\tau\,\bar\nu\,X)/\Gamma\,(\bar B\to e\,\bar\nu\,X)$
as a function of $r=\tan\beta/m_{H^\pm}$.  The shaded area between the
solid lines is our result.  The area between the dashed lines gives the
free quark decay model result.  The dash-dotted lines give the
experimental $1\sigma$ bounds.  }
\end{figure}

\begin{figure}
\caption[b]{
The tau polarization, $A_{\rm pol} =2\tilde\Gamma/\Gamma$, as a function
of $r=\tan\beta/m_{H^\pm}$.  The shaded area between the solid lines is
our result.  The area between the dashed lines gives the free quark
decay model result. }
\end{figure}

}%end tighten (references & figure captions)

\end{document}